# Efficient Search (RES) for One-Hop Destination over Wireless Sensor Networks

[1]Abdul Razaqueand [2]KhaledElleithy

**Wireless and Mobile communication (WMC) Laboratory**
**Department of Computer Science and Engineering**
**University of Bridgeport-06604, USA**
[1]arazaque, [2]elleithy {@bridgeport.edu}

**Abstract: -**The revolution of wireless sensors networks (WSNs) has highly augmented the expectations of people to get the work done efficiently, but there is little bit impediment to deal with deployed nodes in WSNs. The nature of used routing and medium access control (MAC) protocols in WSNs is completely different from wireless adhoc network protocols. Sensor nodes do not have enough capability to synchronize with robust way, in resulting causes of longer delay and waste of energy. In this paper, we deploy efficientenergy consuming sensors and to find one hop robust and efficient destination search in WSNs.

We firstly deploy BT (Bluetooth enabled) sensors, which offer passive and active sensing capability to save energy. This work is a continuation of previous published work in [2]. The BT node is supported with efficient searchmethodss. The main objective of this contribution is to control different types of objects from remote places using cellular phone.

To validate our proposed methodology,simulation is done with network simulator (ns2) to examine the behavior of WSNs. Based on simulation results, we claim that our approach saves 62% energy spent for finding best one-hop destination as compared with existing techniques.

**General Terms:** Design, Experimentation, Performance, Algorithms.
**Key Words and Phrases:**BT sensor, WSNs, preservingenergy, one-hop efficient search.

## 1. INTRODUCTION

The research community have been observing swift scientific and technological advancement since last four decades. The technological progress in compactness of microprocessors has made significant expansionin WSNs [4]. WSN is a fast growing segment attracting people around the world [6].

The goal of WSNsis to facilitate acommunication bridge between the users and the environment using sensors andcomputing devices. WSNs also combine several professions through many application domains, e.g., health, education, security and social care. Many devices are now embedded with computing power like home appliances and portable devices (e.g., microwave ovens, programmable washing machines, robotic hovering machines, mobile phones and PDAs). These devices help and guide us to and from our homes (e.g., fuel consumption, GPS navigation and car suspension) [10]. Ambient Intelligence (AmI) involves compact power that is adapted to achieve specific tasks. This prevalent accessibility of resources builds the technological layer for understanding of WSNs [3].

Information and communications technologies (ICT) have highly been accepted as part of introducing new cost-effective solutions to decrease the cost of energy in WSNs [2]. For example, the use of WSN in home that is equipped with AmI to support people in their homes. These developments have led to the introduction of energy efficient consuming sensors that are capable of monitoring the constraints of humans as well as living environment [5].

Furthermore, expansion in sensors, along with advances of software applications, makes it possible today to control any deviceanytime and everywhere but consumption of energy is an immense problem. These solutions not only improve the quality of education, security and health of people in their own homes but also provide afaster way of communication to interact with devices all over the world [9]. Meanwhile, the exploitation of mobile devices in WSNs provides more flexibility, intelligence and adaptivity to interact with devices dynamically in any environment [12].

Such technological advances make it possible to deploy mobile phones not only as terminal, but also as remote controller for several devices [2, 11]. In this paper, we introduce a novel paradigm to facilitate controlling remotely available servers and different devices using minimum energy consumption.

## 2. PROPOSED ARCHITECTURE

WSNs support more complicated communications between multiple agents without limitations of time. These agents get input and output to facilitate users to achieve potentially multifaceted tasks at high speed [13].The current systems supply some level of adequate service within specific boundaries. The proposed architecture in this paper consists of specific type of BT sensors and mobile phone. It is used to interact with different types of servers and other devices. Initially,the mobile device is deployed to control remote servers and several types of devices through WSN.

The basic architecture of this paradigm is shown in figure 2 that only targets ubiquitous communication. The BT node provides self-directed prototyping platform based on microcontroller and Bluetooth radio. Bluetooth has been introduced as exhibition platform for research to deploy in distributed sensor networks, wireless communication and ad-hoc networks. It is composed of microcontroller, separate radio and ATmega 128. The radio of BT node comprises of two radios: first is low power chipcon CC1000 suited for ISM-band broadcast. It works same as Berkeley MICA2 mote does. This supports BT node to create multi-hop networks. Second radio is Zeevo ZV4002 supported with Bluetooth module.

The BT node provides multiple interfaces to control many devices at the same time but it consumes of lot of energy during the sensing time. To resolve this problem we implementrobust and efficient search at one-hop destination method. We use different protocols and standards in our architecture. Most of sensors do not communicate with Zigbee/ IEEE 802.15.4 standard [2]. Wi-Fi and Bluetooth are also not compatible because both utilize unlicensed 2.4 GHZ ISM bandwidth. So they are using the same bandwidth which causes interference between them. In addition, both are transmitting data at binary phase shift keying (BPSK) and quadric phase shift keying (QPSK). Our selection of BT node sensors is to provide the compatibility with Zigbee, Bluetooth and Wi-Fi. It supports different types of applications and having multitasking support.

Figure 1 shows simple architecture that supports WSN applications. BT node consists of several drivers that are designed with fixed length of buffers. The buffers in BT node are adjusted during compile time to fulfill severe memory requirements. Available drivers include real time clock, UART, memory, I2C, AD converter and power modes.

In this paper we introduce a theoretical framework that facilities compatibility of mobile phoneswith BT node sensors over wireless sensor networks. We install Asus WL-500GP that maintains IEEE 802.11 b/g/n standards that is equipped with USB port to interlink with sensors. We also use Zigbee USB adapter/ IEEE 802.15.4 to provide the communication between sensors. Zigbee/ IEEE 802.15.4 also provide the capability to sensors to maintain multi-hop communication. USB ports and adapters provide promising platform to interlink the mobile phones with sensors for establishment of connectivity to control remote devices. We deploy a highly featured wireless sensor network shown in figure 2.

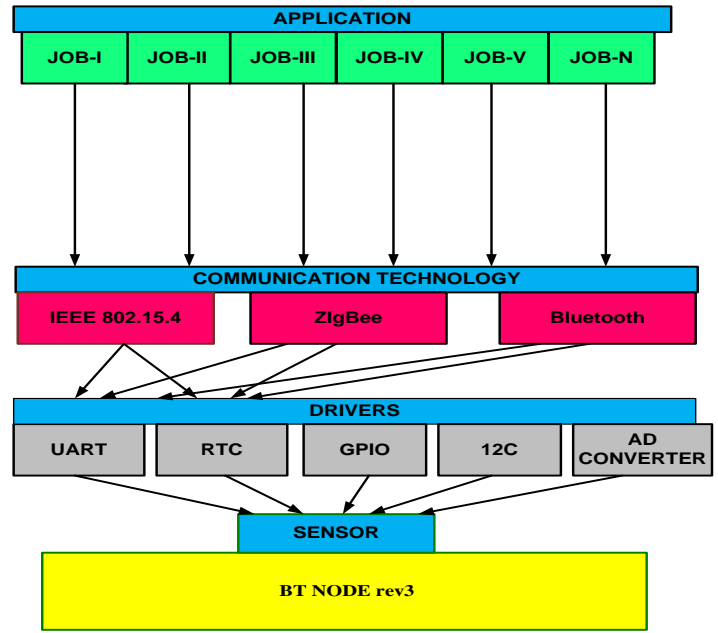

Figure 1: Architecture for Wireless sensor application

The beauty of our WSN is distribution into different regions. Each region has one boundary node that coordinates with boundary node of other regions. The coordination process is also validated with lemma and several definitions discussed in section 3. Participating sensors go automatically into active and passive modes for saving the energy [2]. The working process approach makes WSN to find faster and robust search.

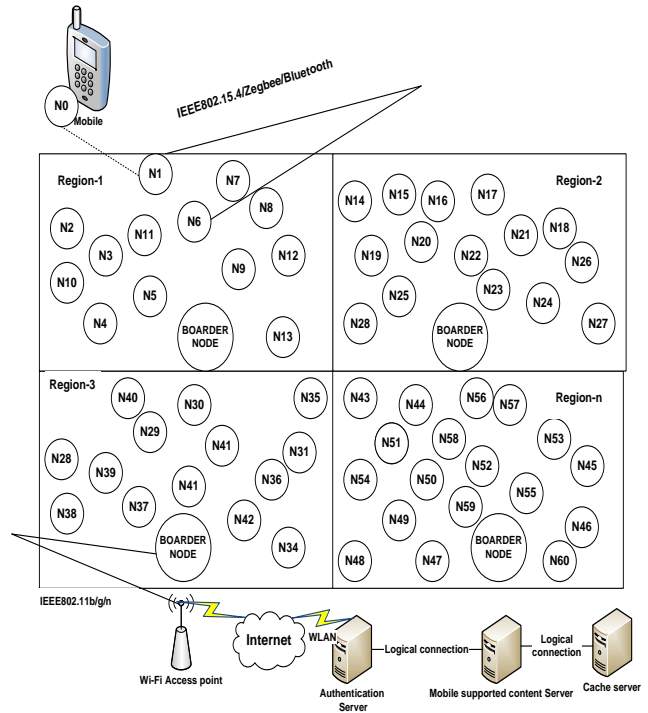

Figure 2: Proposed architecture for WSN

## 3. Efficient Search for one-hop destination

This search is purely based on 1-hop shortest path information and route discovery. The mechanism separates global topology and local connectivity. Suppose that direct graph D = (V, A) is consisting of the set of sensor nodes V. The set of edges are called arcs that are A ⊆ V2. It helps to differentiate between 1-hop destination and more than 1-hop destination nodes. The digraph distance between nodes is simply the number of shortest path between them [7]. We assign name to each sensor node in V. The local route discovery method is based on relay scheme that works as discussed below.

For any destination node in 'V' is specified by name v, the scheme targets the 1-hop destination nodes u on basis of stored information in routing table regarding shortest path 1-hop destination node. Each 1-hop destination node provides the shortest path to its predecessor; finally destination v is obtained with shortest path. We apply the method in [8] for estimation of global technology of sensors by dividing nodes into routable boundaries and extracting adjacency associations between these boundaries. The objective of creating each boundary is to make the topology simpler, so that routing process works efficiently within boundaries. For set of sensor nodes 'V' and communication digraph 'D', we assume D is connected, since we just considered connected components independently. We denote u : (u,v) ∈ A is hop count of the shortest path between u,v in communication digraph.

**Definition 1.** For a digraph D = (V, A), the set of 1-hop destination nodes for vertex v that is explained as Γ−D(v) = {u : (u,v) ∈ A}, and beyond of 1-hop destination nodes are explained as Γ+D(v) = {u : (u, v) ∈ A}. We explain 1-hop destination nodes of a vertex v as union with set of 1-hop destination nodes and the set of more than 1-hop destination nodes, ΓD(v) = Γ+D(v) U Γ−D(v). It has out-degree and in-degree deg+D (v) = |Γ+D (v)| and deg−D (v) = |Γ−D (v)| respectively.

It is clear that degD (v) ≠ deg−D (v) + deg+D (v) and parity do not essentially hold. We may again exclude subscript if digraph DG = (V, A) is clear from context. The weighted graphs also get association of assorted length, cost and strength. We only focus on edge-weighted graph that is opposite to node-weighted graphs. We also need to restrict edge weights to 1 that capitulate an un-weighted graph. Consider digraph DG = (V, A) and its subset for regions R ⊂ V, explain boundary B (v) of a node. Therefore, v ∈ R and whose nearest region is v. It can formally be written as:
B (v) = {u ∈ V | ∀w ∈ R, τ (u, v) ≤ τ (u, v)}

**Lemma1.** For any node u ∈ B (v), the shortest path from node u to v is completely included in B (v).

**Proof:** If lemma were incorrect, there would exist w ≠ B(v) on the shortest path from node u to destination node v. Therefore τ(w,v) < τ(w,u). and such that τ(x,v) ≤ τ(x,w)+ τ(w,v) < τ(x,w) + τ(w,u) = τ(x,u). This statement contradicts with hypothesis such as x ∈B(u); thus lemma must be correct. One inference of this lemma is connection of boundary cells on spanning graph. Boundary cells are dirichlet, connecting all points of sensor field. Boundary has simple topology in all dimensions that is stronger point of connectivity. The simpler topology helps to make subsets of sensor fields, when sensor filed experiences large holes. Thus, edges u1, u2 ϵ B (v).

**Definition 2.** Let PDG (x, y) denote set of paths from x to y in direct graph (DG). Hence, SDG (x, y) is the distance (S) between nodes x, y in DG, which shows shortest path from node x to y. It can be computed as:
$SDG(x,y) = \min l(p) \in p \in PDG(x,y)$ (1)
If SDG (x, y) = ∅ then SDG (x, y) = ∞, Therefore SDG(x, Ė) between node x and subset of nodes Ė ⊆ E that is defined as:
$SDG(x, \dot{E}) =$ (2)
Thus, Ẋ, Ė⊆ E, it is distance between two subsets that can be computed as:
$\min SDG (x, \dot{E}) \& x \in \dot{E}$ (3)
add random infinitesimal if unique path is required.

**Lemma 2.** Let simple path P= (d, a1, a2,…,$a_{e-1}$, t) that connects two boundaries nodes d= a1 and t= $a_e$ with e edges and path of length is l(P). The related boundary path p* has maximum length in boundary dual graph (BG*) such as l(P*) ≤ e * l(P*).

**Proof:** The path includes e-1 more than 1-hop destination nodes and e edges that pass through at most e +1 in the different boundaries of regions. The most of regions e-1 are intrusive regions, it means that original path does not direct through related boundary nodes but shortest boundary path does. The l(P) in original graph is sum of edge weights that can be defined as:

$$l(P) = d(s,t) = \sum_{i=0}^{e-1} w(ai, ai+1)$$ (4)

d* ($B_{bou}(a_i)$, $B_{bou}(a_i+1)$) (5)
It is edge (e) between two nodes of boundaries on path P* that is bounded as follows:
P*= [d* ($B_{bou}(a_i)$, $B_{bou}(a_i+1)$) ≤ d ($B_{bou}(a_i)$, $a_i$) + W ( ($a_i$),

($a_i$+1) + d (($a_i$+1) , $B_{bou}(a_i+1)$)] (6)

From the set boundary of regions, we observe that d ($a_i$, $a_i$ +1) ≤ d($a_i$, $B_{bou}(a_k)$) for all k.

On basis of assumption, we say that s and t are also boundary nodes that could be source and target nodes defined as follows:
d($a_i$, $B_{bou}(a_i)$ ≤ d (s, $a_i$)d($a_i$, $B_{bou}(a_i)$) ≤ d ( $a_{i,}$ t)= d($B_{bou}(a_i)$,

$a_i$) : It yields:

l(P*) ≤ d* (s, t) = d* (s, $B_{bou}(a_1)$)

$$\sum_{i=1}^{e-2}[d(B_{bou\,(a_i)}, a_i + w_{(a_i,a_{i+1})}) + d(a_i + 1, B_{bou(a_{i+1})_{i+1}})] + d*(B_{bou(a_{e-1})}, t) \leq w(s,a_i) + d(a_i, B_{bou(a_i)})$$

(7)

$$\sum_{i=1}^{e-2}[d(B_{bou\,(a_i)}, a_i) + d(a_i + 1, B_{bou(a_i + 1))}]\sum_{i=1}^{e-2}d(a_{i,a_{i+1}}) + d(B_{bou(a_e - 1)}, (a_e - 1, t) \leq$$

$$(s,t) + \sum_{i=1}^{e-2}d(s, a_{i,}) + (a_{i,t,})]$$ (8)

Simplifying the equation (8), we get as:
e. l(P)     (9)

Bound is found to be tight because constructions exist. For example, If any choice for m > ϵ > 0, edge weights of graph should be selected and given as follows:
d($a_i$, $B_{bou(ai)}$) = m — ϵ, w($a_i$, $a_i$+1) = ϵ     (10)

and

w(s, $a_i$) = w($a_{e-1}$,t) = m     (11)

Since 2m + (e—2) ϵ is the length of given path and 2m +(e-2) ϵ 2(e-1)*(e- ϵ) is the length of boundary path.

Therefore, the worst case for 'ϵ' can be written as:
Є → 0, and ratio can be shown as follows:

$$\frac{l(P*)}{l(P)} \to e$$ (12)

If boundary nodes are available on shortest path, thus maximum expansion is assured to be shorter than number of edges on shortest path. We hereby prove that maximum expansion is proportional to largest gap between boundary nodes on path.

**Lemma 3:** For each region 'R', the flooding message ($R_n$, $f_m$) =1 can provide the shortest distance region to each node in Sensor network N.

**Proof:** Here $R_n$ is number of regions and $f_m$ is flooding message that can be sent from one region to other region. Thus, each node v maintains the list of current shortest path region ($R_v$) with shortest distance ($D_v$)in network. Assume $R_v$ =∅ and $D_v$ =∞. Therefore, on obtaining flooding message in any region ($R_n$, $f_m$), we apply three conditions as follows:

**Condition-I:**
if$f_m$>$D_v$[messageis discarded]

**Condition-II:**
if$f_m$= $D_v$, then we apply two cases:
if$R_n$∈$R_v$[messageis discarded]
  A. if  $R_n$∉$R_v$;then
$R_n$ + $R_v$and [Broadcasted message] to all the neighbor nodes.( $L_n$ , $f_m$ +1) [Broadcasted message]

**Condition-III:**
if $f_m$<$D_v$; ; then$R_v$ = ($R_n$) , $D_v$= $f_m$and( $R_n$ , $f_m$ +1) [Broadcasted message] to all the neighbor nodes.

If regions initiate the process at same time, every message travels at the same speed. Thus, $f_m$ messageof any region is dropped when it begins to penetrate in the boundaries of other cells. This causes of cutting down the transmitted messages in flood. At the end of process, the list $R_v$holds set of regions which are at the shortest distance to node v. At this point, each node knows its boundary of region in sensor network N. Hence, the flooding message ($R_n$, $f_m$) =1 provides the shortest path to node.

## 4. SIMULATION AND ANALYSIS OF RESULT

Real wireless sensor environments use low power radios and are known due to high asymmetrical communication range and stochastic link attributes. The simulation results could be different from realistic experimental results [1]. If network simulator makes only simple assumptions on wireless radio propagation, exact simulation with features of real wireless radios and diverse transmission powers can be significant. The WSN is distributed into different regions as illustrated in figure 2 to make the sensors more convenient to collect information quicker.

We have already discussed about boundary node that is playing role as anchor point (AP) or head node. We have set one boundary node in each region. Boundary node forwards the collected information of its region to next region. In our case, it is not necessary that boundary node may always coordinate with only boundary node of other region but it can forward the gathered information at 1-hop destination either boundary node or common node.

We have simulated several types of different scenarios by increasing the number of sensors.These scenarios are real test of WSN. We have deployed 35 to 140 sensors within network area of 160m × 160m. Area is divided into 40m x 40m regions. Sensors are randomly located within each region.

The sink in each scenario is located at (140, 60), but bandwidth of node is 50 Kbps and maximum power consumption for each sensor is set 160 mW, 12 mW and 0.5 mW for communication, sensing and idle modes respectively but in our case, there is no idle mode. Sensors either go to active or sleep mode. Each sensor is capable of broadcasting the data at 10 power intensity ranging from -20 dBm to 12 dBm. Total simulation time is 140 minutes and there is no pause time during the simulation but we have set 30 seconds for initialization of phase at beginning of simulation. During this phase, only sensors onto sink remain active and remaining sensors of all regions go into power saving mode automatically. We have chosen well known energy saving approaches. We have specially checked the performance of existing approaches at routing level. These existing approaches include minimum energy relay routing (MERR), minimum-transmission-energy (MTE), direct Transmission (DT) and optimal routing (OR).The results shown in this section are average of 10 simulation runs.

## 4.1 Efficiency of proposed WSN

To valid this environment of WSN for handling several devices, we conduct several tests from different perspectives. Having presented the mathematical models, we now evaluate the efficiency of WSN. We present the amount of energyused in the network for a number of sensors as shownin Figures t to Figure 6

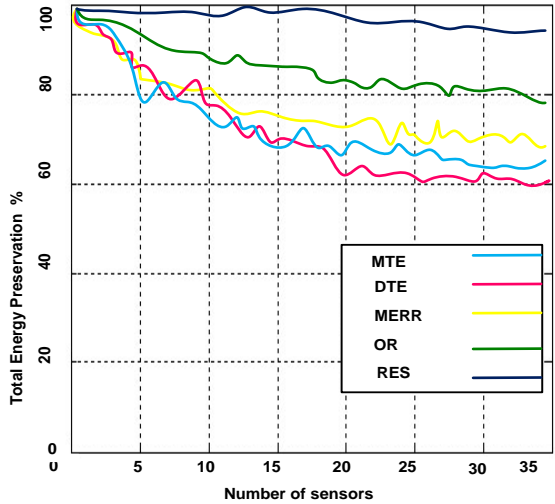

Figure 3: Preservation of energy in % with 35 sensors

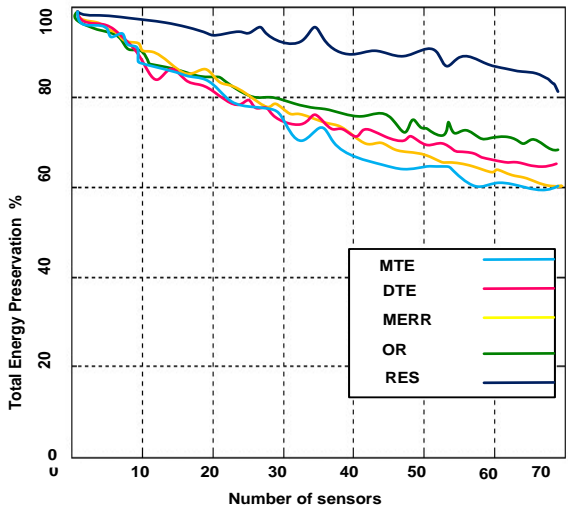

Figure 4: Preservation of energy in % with70sensors

Our simulated network shows that our proposed paradigm achieves almost 100% efficiency and saves 62% energy using maximum 140 sensors. We establish 15 sessions simultaneously in order to determine the actual behavior of the network in highly congested environment. If we have less number of sensors, it is hard to establish many sessions at the same time.It is very noticeable direction of this research that 140 sensors can provide path-connectivity for 15 mobile devices to interact with remotely placed devices at same time. In addition, one mobile node can interact with multiple devices at same time. Question is why to deploy more sensors in that area? The answer is the availability of several servers and devices at the different places.

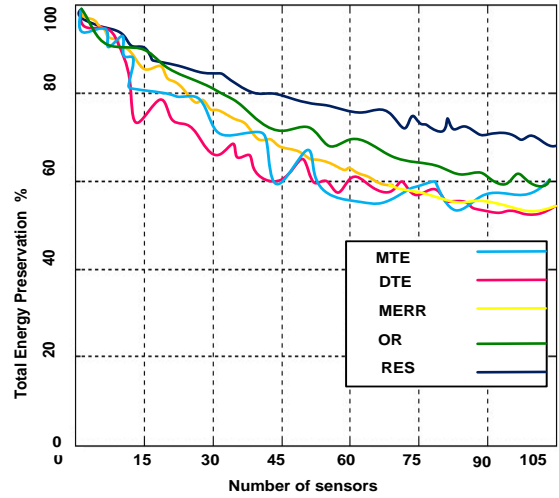

Figure 5: Preservation of energy in % with105sensors

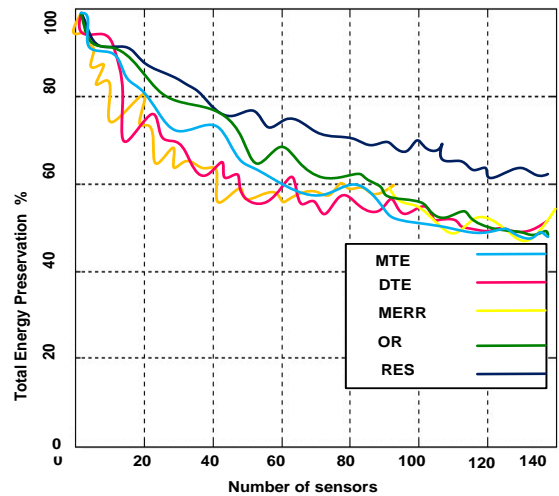

Figure 6: Preservation of energy in % with140sensors

More sensors are required to find path and provide the connectivity for enough number of mobile devices working concurrently. Figure 7 shows the trend of network during the time of 140 minutes of simulation at the same number of established sessions. During this duration, the mobile devices get 99.2% coverage of network whereas existing network affects the performance of network as time increases of communication. This is another weakness of existingWSNs. Our proposed method produces stable efficiency during all simulation time. It is proved that duration of simulation either increases or decreases; do not affect the efficiency in our case. The minimum number of sensors required for covering area can be calculated as follows:

$N_{min(s)} = \frac{2\pi * A}{3\pi R^2}$; Where $N_{min(s)}$ is minimum number of sensors to cover whole area to maintain connectivity and coverage. 'r' is sensing range of sensor. It should be assumed that sensing range is smaller than dimensions of monitoring area. $\frac{N_{min(s)}}{N_{max(s)}}$ shows maximum number of sensors. 'R' is distance of total network. We prove this with help of lemma4.

**Lemma 4:** $\frac{N_{min(s)}}{N_{max(s)}}$ is upper bound on R and $N_{min(s)}$ is lower bound on $S_i$, where $N_{min(s)} = \frac{2\pi * A}{3\pi R^2}$.

**Proof:** Let upper bound be linear on R with maximum number of sensors (total number of sensors) $N_{max(s)}$ whereas lower bound on $S_i$ is invariant with $N_{max(s)}$. In addition, these bounds are not considered tight as long as they do not consider transmission radius '$T_r$' of sensors. However, we need better heuristic solution to follow these bounds closely if irrespective of changes occur in the parameters of network. Hence, the lifetime of the network should be linearly with $N_{max(s)}$, and Si to be constant with $N_{max(s)}$.

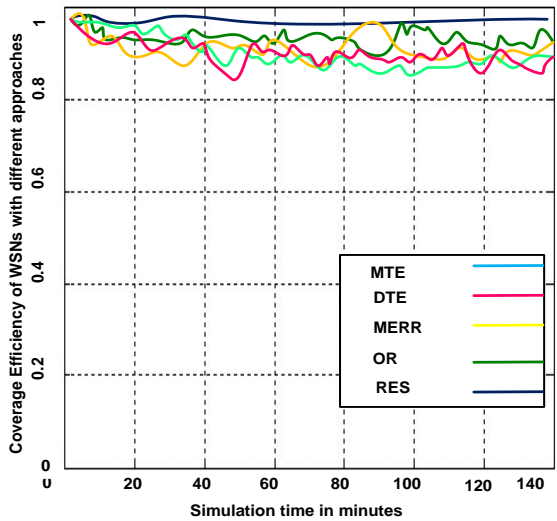

Figure 7. Coverage efficiency of network at different intervals

## 5. CONCLUSION

In this paper, we have introduce anefficient search algorithmfor one-hop destination to save energy that provides access to control remotely available servers and several types of devices through mobile cell. This unique type of research deploys BT sensors in wireless network to control the devices. Furthermore, WSN is divided into number of N-regions. Each region consists of one boundary node that is static and is responsible to communicate with other regionsfor saving energy. When the sensor finishes its assigned task, it automatically goes to sleep mode. To validate the proposed WSN, we havesimulated the network using ns2.35-RC7. On the basis of simulation results, we prove that our proposed research saves maximum amount of energy as comparedtoexisting WSNs of saving energy. In addition, we have achieved the objective ofcontrolling the devices from remote places by consuming minimum energy resources. In the future, we are planningto implement this simulation based network into testbed to control several devices simultaneously.